\documentclass[a4paper,fleqn]{cas-dc}
\usepackage[numbers]{natbib}
\usepackage[english]{babel}
\usepackage[utf8]{inputenc}
\usepackage[T1]{fontenc}
\usepackage{amsthm}
\usepackage{amssymb}
\usepackage{bm}
\usepackage{color}
\def\tsc#1{\csdef{#1}{\textsc{\lowercase{#1}}\xspace}}
\tsc{WGM}
\tsc{QE}
\tsc{EP}
\tsc{PMS}
\tsc{BEC}
\tsc{DE}
\newcommand{\dbar}{d\hspace*{-0.08em}\bar{}\hspace*{0.1em}}
\begin{document}

\title [mode = title]{Quantum heat machines enabled by the electronic effective mass}                    
\tnotetext[1]{This document is the results of the research
   project funded by the CNPq(Brazilian Agency).}
\author[1]{Cleverson Filgueiras}[                        orcid=0000-0002-3459-8894]

\ead{cleverson.filgueiras@ufla.br}
\address[1]{Departamento de F\'{i}sica, Universidade Federal de Lavras, Caixa Postal 3037, 37200-000 Lavras-MG, Brazil }
\fntext[1]{Also at Departamento de F\'{i}sica, Universidade Federal da Para\'{i}ba, Caixa Postal 5008, 58059-970, João Pessoa, PB, Brazil.}


\begin{abstract}
In this letter, we analyze a conceptual design for the operation of an Otto cycle heat machine driven by adiabatic modifications on the electronic effective mass. Such tailoring of it can be implemented, for instance, via the application of external electron fields in some materials, as in Gallium Nitride (GaN). We show that due both the energy quantization on this structure and the adiabatic transformation of the effective mass, the machine performance can be improved. The realization of classically inconceivable Otto machines, with an incompressible working substance, can be realized as well. Our finds hold in cases where the electronic effective mass, in the low temperature regime, remains constant during the isochoric strokes. 
\end{abstract}
\maketitle
\section{\label{sec:level1}Introduction}

In the field of {\it quantum thermodynamics}\cite{doi:10.1080/00107514.2016.1201896}, an active branch of research is the investigation of quantum heat engines(QHE), the devices that convert the heat energy into work by means of a quantum system as their working substance \cite{first}.  This topic has been attracting a lot of interest over the last few years.  From the fundamental point of view, the conceptual designs of such engines are proposed\cite{e18050173}. Here, we intend yet to propose a conceptual design of such a QHE, for electronic systems. As it is well known, the carrier {\it effective mass} has an important role in electronic systems\cite{kittel2005introduction}. It is parameter that can be tailored in some materials by applying electric fields\cite{tailoring,silverinha}. The Gallium Nitride (GaN) supports higher critical electrical field which makes it very attractive for power semiconductor devices. The electronic effective mass of GaN can be adjusted by the application of electric fields across it and its exact value depends on the electron density and scattering time\cite{PhysRevB.86.161104}. 

 In this paper we will examine an Otto engine operated with an ideal electron gas contained in an one-dimensional (1D) infinite well potential. The problem is modeled by a single electron with an effective mass as, for example, that in GaN. We consider the adiabatic application of the electron fields, tailoring the effective mass which in turns changes the energies of select quantum states. In what follows, we will consider the electrons confined in such a structure considering that the electron effective mass keeps its value in the low temperature regime\cite{temperature}.  Therefore, we will consider the electronic effective mass constant during the isochoric strokes. 
 
\section{\label{sec:level2}The quantum Otto cycle for an electron in a box}
\begin{figure}
\begin{center}
\includegraphics[width=0.5\textwidth]{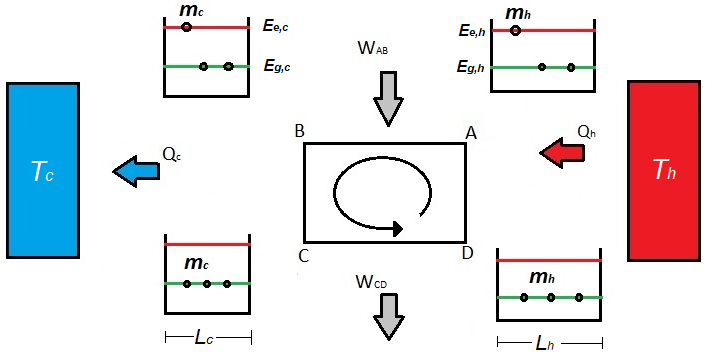}
\caption{\small A Quantum Otto Cycle for an electron system in an infinity box. The adiabatic strokes are consisted in modifications in the electronic effective mass by the application of an external electrical field. In the isochoric ones, the electronic effective mass remains constant provided $T_c$ and $T_h$ are kept in the low temperature regime.} \label{Mass}
\end{center}
\end{figure}

\begin{figure}
\includegraphics[width=0.5\textwidth]{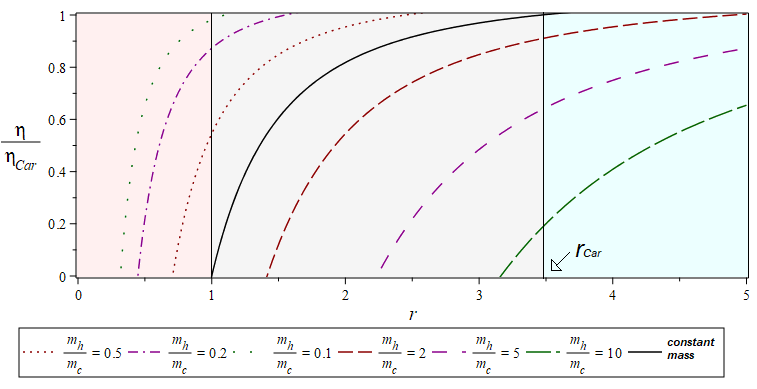}
\caption {\small Efficiency normalized to Carnot efficiency versus the compression ratio, $r$, for fixed values of $\frac{m_h}{m_c}$. We considered $\gamma=3$ and $\frac{T_h}{T_c}=12$. In this case, $r_{Car}=3.464101615.$ Notice that the adiabatic changes in the electronic effective mass lead to existence of engines in regions where the work extraction is not expected in the constant mass case.} \label{eff}
\end{figure}

The Otto cycle, depicted in Fig. \ref{Mass},  consists in four strokes that connect
different states of the system, A, B, C and D.``c''(``h'') labels cold(hot). In the first stage, the electron (working substance) with an effective mass $m_h$ is coupled to a hot bath of temperature $T_h$. The Hamiltonian is $H_h = \frac{P^2}{2m_h}$, where
$P$ is the momentum operator. Some heat is transferred from the heat bath to the working substance. The infinitesimal heat transferred can be written as $\dbar Q_h =\sum_n E_{n,h} dp_{n,h}$, with $n=1,2,...$ During this process, the occupation probabilities, $p_{n,h}$(given in terms of the Fermi–Dirac statistics in the canonical ensemble), are modified and each eigenenergy is kept fixed at $E_{n,h}$. In A, the occupation probability of the $n^{th}$-level is $p_{n,A}$. The size of the system is $L_h$. In the second stage, the working substance is isolated from the hot bath and undergoes an adiabatic process in which the effective mass is adiabatically modified until it reaches a value $m_c$ at B. The Hamiltonian at $B$ is $H_c =\frac{P^2}{2m_c}$. By keeping the process adiabatic, it ensures that the level populations do not change, according to the quantum adiabatic theorem\cite{AdiabaticTheorem}. Then, $p_{n,A} = p_{n,B}$. The change in the energy of the system can be attributed solely to work, which we call $W_{AB}$. Its general infinitesimal form is $\dbar W=\sum_n p_n dE_n$\cite{PhysRevE.76.031105}. Next, the system is coupled to a cold thermal bath at temperature $T_c$ and it reaches thermal equilibrium at C. Thus, the occupation probability for the $n^{th}$-level eigenenergy of $H_c$  at $C$ is written as $p_{n,C}$. The size of the system does not change and the change in the energy of the system can be attributed to heat exchange with the cold bath, $\dbar Q_c =\sum_n E_{n,c} dp_{n,c}$. In the following stroke, the system is decoupled from the cold bath and the electric field is adiabatically changed, returning the size of the system to $L_h$ and the effective mass to the value $m_h$ at D. The level populations do not change, so we have $p_{n,D} = p_{n,C}$. Therefore, all the energy exchanged is work, which we call $W_{CD}$. The thermodynamic cycle is then, closed. The heat exchanged with the baths is given by the energy difference between the initial and the final states of
the isochoric strokes. Then, 
\begin{eqnarray}
Q_h=\left<H_h\right>_A-\left<H_h\right>_D=\sum_n E_{n,h}\left[p_n(A)-p_n(C)\right]\;
\end{eqnarray}
and
\begin{eqnarray}
Q_c=\left<H_c\right>_C-\left<H_c\right>_B=\sum_n E_{n,c}\left[p_n(C)-p_n(A)\right]\;.
\end{eqnarray}
By energy conservation, the net work after completing a cycle will be given by
\begin{eqnarray}
W&=&-Q_h-D_c\;.\label{work1}
\end{eqnarray}
Considering the case where only the first two levels are populated( $E_1\equiv E_{g,i}$ and $E_{2,i}\equiv E_e$, $i=,h,c$), the Eq. \ref{work1} yields the efficiency as\cite{PhysRevLett.120.170601}
\begin{eqnarray}
\eta=-\frac{W}{Q_h}=1-\frac{\Delta_c}{\Delta_h}=1-\frac{m_h}{m_c}\left(\frac{L_h}{L_c}\right)^2\;.\label{eta}
\end{eqnarray}
In order to derive Eq. (\ref{eta}), we considered an electron in an 1D infinity box, so $E_n=\frac{\hbar^2\pi^2n^2}{2mL^2}$, $n=1,2,3,..$. We wrote $\Delta_c=E_{e,c}-E_{g,c}$ and $\Delta_h=E_{e,h}-E_{g,h}$. We have also used $p_1(A)+p_2(A)=1$ and $p_1(C)+p_2(C)=1$. The condition for work extraction, $W < 0$, is $\frac{T_h}{T_c}>\frac{\Delta_h}{\Delta_c}>1.$ For a two-level system, the probabilities are canceled out in it. In the Fig. \ref{eff}, we plot the efficiency normalized to the Carnot efficiency versus the compression ratio, $r\equiv\frac{L_c}{L_h}$, for fixed values of $\frac{m_h}{m_c}$.  This last one is given by $\eta_{Car}=1-\frac{T_c}{T_h}=1-\frac{1}{r_{Car}^{\gamma-1}}\;,$
where $r_{Car}$ is the compression ratio and $\gamma=\frac{C_p}{C_v}$ is the ratio of specific heats at constant pressure and at constant volume.
By tailoring the effective mass, we observe the engine operation at Carnot efficiency and it can be modified in comparison to the common Otto machine(constant mass). A quantum enhanced performance is observed for $m_h<m_c$ and it can be traced to the change in the relation between temperature and the population distribution for the adiabatic modifications in the effective mass due to the application of electric fields. The efficiency is reduced otherwise. We emphasize that this analysis relies only on energy quantization and the constant level populations in adiabatic strokes. The ratio, $\frac{m_h}{m_c}$, could be optimized for reaching maximum work extraction at any compression ratio. The realization of an Otto machine with an incompressible working substance, for which $r=1$, can be realized. The efficiency at maximum power output, for instance, should be latter addressed, at first, in the framework of {\it endoreversible thermodynamichs}\cite{e20110875}.  
\section{Concluding Remarks}
In this contribution, we addressed an Otto cycle heat machine driven by an electron gas in a 1D infinity box. By considering the adiabatic transformation of its effective mass, we showed that the machine performance can be modified. The realization of Otto machines with an incompressible working substance, $r=1$, is possible to be implemented in such an electronic system. By optimizing the ratio $\frac{m_h}{m_c}$, a maximum work extraction at any compression ratio could be reached. This work was intended study a concept rather than the implementation of a practical protocol. We believe that experimentally it would be easier to
control the intensity of an applied electric field in the quantum well made up with GaN. The change of temperature regime can be investigated elsewhere. 

\bibliographystyle{model1-num-names}
\bibliography{cas-refs.bib}

\begin{thebibliography}{12}
\expandafter\ifx\csname natexlab\endcsname\relax\def\natexlab#1{#1}\fi
\providecommand{\bibinfo}[2]{#2}
\ifx\xfnm\relax \def\xfnm[#1]{\unskip,\space#1}\fi
\bibitem[{Vinjanampathy and Anders(2016)}]{doi:10.1080/00107514.2016.1201896}
\bibinfo{author}{S.~Vinjanampathy}, \bibinfo{author}{J.~Anders},
\newblock \bibinfo{journal}{Contemporary Physics} \bibinfo{volume}{57}
  (\bibinfo{year}{2016}) \bibinfo{pages}{545--579}.
\bibitem[{Scovil and Schulz-DuBois(1959)}]{first}
\bibinfo{author}{H.~E.~D. Scovil}, \bibinfo{author}{E.~O. Schulz-DuBois},
\newblock \bibinfo{journal}{Phys. Rev. Lett.} \bibinfo{volume}{2}
  (\bibinfo{year}{1959}) \bibinfo{pages}{262--263}.
\bibitem[{Muñoz et~al.(2016)Muñoz, Peña, and González}]{e18050173}
\bibinfo{author}{E.~Muñoz}, \bibinfo{author}{F.~J. Peña},
  \bibinfo{author}{A.~González},
\newblock \bibinfo{journal}{Entropy} \bibinfo{volume}{18}
  (\bibinfo{year}{2016}) \bibinfo{pages}{173}.
\bibitem[{Kittel(2005)}]{kittel2005introduction}
\bibinfo{author}{C.~Kittel}, \bibinfo{title}{Introduction to solid state
  physics}, \bibinfo{publisher}{Wiley}, \bibinfo{edition}{8th} edition,
  \bibinfo{year}{2005}.
\bibitem[{Naveh and Laikhtman(1995)}]{tailoring}
\bibinfo{author}{Y.~Naveh}, \bibinfo{author}{B.~Laikhtman},
\newblock \bibinfo{journal}{Applied Physics Letters} \bibinfo{volume}{66}
  (\bibinfo{year}{1995}) \bibinfo{pages}{1980--1982}.
\bibitem[{Silveirinha and Engheta(2012{\natexlab{a}})}]{silverinha}
\bibinfo{author}{M.~G. Silveirinha}, \bibinfo{author}{N.~Engheta},
\newblock \bibinfo{journal}{Phys. Rev. B} \bibinfo{volume}{86}
  (\bibinfo{year}{2012}{\natexlab{a}}) \bibinfo{pages}{245302}.
\bibitem[{Silveirinha and Engheta(2012{\natexlab{b}})}]{PhysRevB.86.161104}
\bibinfo{author}{M.~G. Silveirinha}, \bibinfo{author}{N.~Engheta},
\newblock \bibinfo{journal}{Phys. Rev. B} \bibinfo{volume}{86}
  (\bibinfo{year}{2012}{\natexlab{b}}) \bibinfo{pages}{161104}.
\bibitem[{Hofmann et~al.(2012)Hofmann, Kühne, Schöche, Chen, Forsberg,
  Janzén, Ben~Sedrine, Herzinger, Woollam, Schubert, and
  Darakchieva}]{temperature}
\bibinfo{author}{T.~Hofmann}, \bibinfo{author}{P.~Kühne},
  \bibinfo{author}{S.~Schöche}, \bibinfo{author}{J.-T. Chen},
  \bibinfo{author}{U.~Forsberg}, \bibinfo{author}{E.~Janzén},
  \bibinfo{author}{N.~Ben~Sedrine}, \bibinfo{author}{C.~M. Herzinger},
  \bibinfo{author}{J.~A. Woollam}, \bibinfo{author}{M.~Schubert},
  \bibinfo{author}{V.~Darakchieva},
\newblock \bibinfo{journal}{Applied Physics Letters} \bibinfo{volume}{101}
  (\bibinfo{year}{2012}) \bibinfo{pages}{192102}.
\bibitem[{Kato(1950)}]{AdiabaticTheorem}
\bibinfo{author}{T.~Kato},
\newblock \bibinfo{journal}{Journal of the Physical Society of Japan}
  \bibinfo{volume}{5} (\bibinfo{year}{1950}) \bibinfo{pages}{435--439}.
\bibitem[{Quan et~al.(2007)Quan, Liu, Sun, and Nori}]{PhysRevE.76.031105}
\bibinfo{author}{H.~T. Quan}, \bibinfo{author}{Y.-x. Liu},
  \bibinfo{author}{C.~P. Sun}, \bibinfo{author}{F.~Nori},
\newblock \bibinfo{journal}{Phys. Rev. E} \bibinfo{volume}{76}
  (\bibinfo{year}{2007}) \bibinfo{pages}{031105}.
\bibitem[{Gelbwaser-Klimovsky et~al.(2018)Gelbwaser-Klimovsky, Bylinskii,
  Gangloff, Islam, Aspuru-Guzik, and Vuletic}]{PhysRevLett.120.170601}
\bibinfo{author}{D.~Gelbwaser-Klimovsky}, \bibinfo{author}{A.~Bylinskii},
  \bibinfo{author}{D.~Gangloff}, \bibinfo{author}{R.~Islam},
  \bibinfo{author}{A.~Aspuru-Guzik}, \bibinfo{author}{V.~Vuletic},
\newblock \bibinfo{journal}{Phys. Rev. Lett.} \bibinfo{volume}{120}
  (\bibinfo{year}{2018}) \bibinfo{pages}{170601}.
\bibitem[{Deffner(2018)}]{e20110875}
\bibinfo{author}{S.~Deffner},
\newblock \bibinfo{journal}{Entropy} \bibinfo{volume}{20}
  (\bibinfo{year}{2018}) \bibinfo{pages}{875}.

\end{thebibliography}
\end{document}